\documentstyle[prl,aps,epsf,multicol]{revtex}
\begin{document}
\draft

\title{Persistence in $q$-state Potts model: A Mean-Field approach}

\author{G. Manoj}

\address{ Department of Physics and Center for Stochastic Processes in Science and Engineering,\\ 
Virginia Polytechnic Institute and State University, Blacksburg, 
VA 24061, USA.}

\date{\today}

\maketitle

\begin{abstract}

We study the Persistence properties of the $T=0$ coarsening dynamics of one dimensional
$q$-state Potts model using a modified mean-field approximation (MMFA). In this approximation,
the spatial correlations between the interfaces separating spins with different
Potts states is ignored, but the correct time dependence of the mean density $P(t)$
of persistent spins is imposed. For this model, it is known that $P(t)$ follows a power-law
decay with time, $P(t)\sim t^{-\theta(q)}$ where $\theta(q)$ is the $q$-dependent
persistence exponent.
We study the spatial structure of the persistent region within the 
MMFA. We show that the persistent site pair correlation function $P_{2}(r,t)$ has the
scaling form $P_{2}(r,t)=P(t)^{2}f(r/t^{\frac{1}{2}})$ for all values of 
the persistence exponent $\theta(q)$.
The scaling function has the limiting behaviour
$f(x)\sim x^{-2\theta}$ ($x\ll 1$) and $f(x)\to 1$ ($x\gg 1$). 
We then show within the Independent Interval Approximation (IIA) that 
the distribution $n(k,t)$ of separation $k$ between two consecutive persistent spins
at time $t$ has the asymptotic scaling form $n(k,t)=t^{-2\phi}g(t,\frac{k}{t^{\phi}})$ where the
dynamical exponent has the form $\phi$=max($\frac{1}{2},\theta$). The behaviour of the scaling
function for large and small values of the arguments is found analytically. We find that 
for small separations $k\ll t^{\phi}, n(k,t)\sim P(t)k^{-\tau}$ where $\tau$=max($2(1-\theta),2\theta$),
while for large separations $k\gg t^{\phi}$, $g(t,x)$ decays exponentially with $x$.
The unusual dynamical scaling form and the behaviour of the scaling function is supported by
numerical simulations.

\end{abstract}


\vspace{1cm}
\begin{multicols}{2}
\section{Introduction}

In recent times, the notions of Persistence and the associated Persistence exponent
has become one of the active topics of research in Non-equilibrium Physics\cite{SATYA}.
In general, the persistence probability $P(t)$ is the probability that a stochastic
variable $\phi(t)$ remains above or below a certain arbitrary value (say, its initial value)
for a time interval $[0:t]$. The idea of Persistence is particularly relevant 
in coarsening systems where $P(t)$ has the natural interpretation of being the fraction
of space in the system which remains in the same phase, starting from a random initial
configuration. In these systems (as in some other systems also) $P(t)$ typically
decays as a power-law at large $t$, with an exponent that is non-trivial to compute
and does not appear to be simply related to other known exponents that characterize the
process.

The coarsening dynamics of the one dimensional $q$-state Potts model at 
zero temperature is one of the few cases where the Persistence exponent $\theta(q)$ is 
known exactly. The solution  was provided by Derrida {\it et.al} through a mapping
of the process to a coagulation model in steady state\cite{DERRIDA}. It was shown that at late times $t$, well beyond
the time scale of equilibration, the fraction of persistent spins left in a finite system
of linear dimension $L$ scale as $P_{q}(t\gg L^{2}, L)\sim L^{-2\theta(q)}$, 
where $\theta(q)$ is given by the non-trivial expression

\begin{equation}
\theta(q)=-\frac{1}{8}+\frac{2}{\pi^2}\left[cos^{-1}\left(\frac{2-q}{q\sqrt{2}}\right)\right]^{2}.
\label{eq:EQ1}
\end{equation}

For times $t\ll L^{2}$, it follows that $P_{q}(t)\sim t^{-\theta(q)}$. An interesting question in this
context is about the spatial distribution of spins which are persistent up to any given time
$t$. Clearly, in a many-body process like the time evolution of Potts model, the probability
that a given spin is persistent is closely linked to the state of other spins. This inter-dependence
of spins is crucial in that it makes the time evolution at any single site strongly
non-Markovian, which makes the computation of the Persistence exponent highly non-trivial. 
A study of the spatial aspects of the Persistence problem is therefore important from the point
of view developing an intuitive understanding of the phenomenon, and also illuminates
the interplay between Persistence decay and the underlying domain coarsening process.

The spatial distribution of persistent sites and its time evolution have been studied
previously through numerical simulations in one dimensional
diffusion equation\cite{ZANETTE}, $q$-state Potts model\cite{MANOJ0,MANOJ1,MANOJ2,STEVE1} ,
two dimensional Ising model\cite{JAIN} and one dimensional Ising model with parallel dynamics\cite{PURU}. 
An analytic study using a rate equation approach under the Independent Interval Approximation(IIA)
has been carried out for one dimensional $A+A\to\emptyset$ model which is closely
related to 1D Ising model\cite{MANOJ2}. It is now generally understood from physical arguments and 
simulations that for a coarsening process
in $d$ dimensions where characteristic length scale has the power-law growth
$L(t)\sim t^{1/z}$, the set of persistent sites forms a fractal structure with
fractal dimension $d_{f}=d-z\theta$ over length scales $r\ll t^{1/z}$\cite{STEVE1,MANOJ3}. The distribution is
homogeneous beyond this length scale. Furthermore, since $d_{f}\geq 0$, the distribution is
expected to be homogeneous over all length scales if $\theta > d/z$. This has important
consequences for systems like the Potts model where $\theta$ changes with the Potts state $q$.
In particular, for Potts model in $d=1$, Bray and O'Donoghue\cite{STEVE1} argued that 
a transition from fractal to homogeneous distribution occurs as $\theta$ crosses $\frac{1}{2}$.
This transition is also marked by an abrupt
change in the dynamical exponent characterizing the separation between persistent sites. 
The characteristic length scale  was conjectured to have the unusual dynamical scaling
form ${\cal L}(t)\sim t^{\phi}$ with $\phi=max(\frac{1}{2},\theta)$.
This conjecture based on physical arguments was supported by numerical simulations\cite{STEVE1}. 

In this paper, we use a Mean-Field approach to address the problem of spatial
distribution of persistent sites in $q$-state Potts model. 
The essential idea behind this approach is as follows. 
It is well-known that the $T=0$ coarsening dynamics of the $q$-state Potts model 
can be mapped to a reaction-diffusion process. In this process, the interfaces between different
species of Potts spins are represented by diffusing particles $A$, which annihilate or coagulate
upon meeting with $q$-dependent probabilities. In the
mean field approach, these diffusing particles are treated as homogeneously distributed, with (time-dependent) 
density equal to the average density in the original reaction-diffusion problem. This approach has been
discussed in some earlier works as a  heuristic argument \cite{CARDY} and as 
a toy model for persistence \cite{STEVE2}.  We argue that this approach
yields a lower bound for the Persistence exponent in the Potts model. We then construct an
artificial model which is devoid of spatial correlations among diffusing particles, but with persistence
exponent tuned to be exactly equal to the Potts model value. We refer to this model as the
Modified Mean-Field Approximation (MMFA) and use this approximation to study the spatial
distribution of persistent sites in $q$-state Potts model. 

We outline our main results at this point. Within the MMFA, we show analytically
that the correlation length for the spatial distribution of persistent spins scales
as $\xi(t)\sim t^{\frac{1}{2}}$ and the equal time pair correlation function
$P_{2}({\bf r},t)$ (defined as the probability
that the spin at origin and the point ${\bf r}$ are persistent at time $t$) has the scaling
form $P_{2}({\bf r},t)=P(t)^{2}f(r/\sqrt{t})$ for any value of $\theta$. This shows that the persistent
spins have a fractal distribution with $d_{f}=1-2\theta$ over length scales $r\ll t^{\frac{1}{2}}$
when $\theta < \frac{1}{2}$, but $d_{f}=0$ when $\theta\geq \frac{1}{2}$. 
We find that the characteristic length scale of the spatial distribution of persistent spins has 
the unusual scaling form ${\cal L}(t)\sim t^{\phi}$ where $\phi$=max($\frac{1}{2},\theta$), in
agreement with the conjecture in \cite{STEVE1}. 
The empty interval
distribution itself has the scaling form $n(k,t)={\cal L}(t)^{-2}g[t,k/{\cal L}(t)]$, where
$g(t,x)\sim t^{-\psi}x^{-\tau}$ for $x\ll 1$ and decays exponentially with $x$ when $x\gg 1$.
The exponents $\psi$ and $\tau$ depends on $\theta$ through the relations $\psi=\theta (2\theta-1)H(\theta-
\frac{1}{2})$ and $\tau$=max[$2(1-\theta),2\theta$], where $H(x)$ is the Heaviside step function.
We support these results with numerical simulations.

The paper is arranged as follows. In the next section, we outline the mean field
approach. In Section III, we introduce the MMFA and compute the pair correlation and Empty
Interval Distribution of persistent sites to characterize their spatial distribution. These
predictions are compared with the results of numerical simulations
in the $q$-state Potts model in Sec. IV. We summarize our results and present our
conclusions in Sec. V. 

\section{The Mean-Field approximation}

In the zero temperature coarsening dynamics of $q$-state Potts model in $d=1$, 
the interfacial points between different species of Potts spins perform independent random
walks on the lattice and annihilate each other with probability $\frac{1}{q-1}$
($\frac{q-2}{q-1}$), or coagulate with probability $\frac{q-2}{q-1}$. In the process,
persistent sites are `wiped out', and the surviving random walkers build up spatial
correlations among themselves. The distribution of intervals between the surviving random
walkers at any (sufficiently late) time $t$ is described by a (stationary) scaling function
which is known exactly for all values of $q$\cite{DERRIDA2}. The average density $n(t)$ at time $t$ decays as

\begin{equation}
n(t)\simeq \frac{q-1}{q}\frac{1}{\sqrt{2\pi t}}
\label{eq:EQ2}
\end{equation}

asymptotically\cite{DERRIDA2}. The essential idea behind the mean field (MF) approximation is to treat the 
random walkers
as forming a homogeneous background of average density $n(t)$, as far as the persistent sites
are concerned. We define the Persistence probability $P(t)$ as 
the probability that the site at the origin is 
unvisited by any walker till time $t$. Then, the probability that the site 
the probability that the site is visited by a walker between time
$t$ and $t+dt$ is $-\frac{\partial P(t)}{\partial t}$.
Let $R(x,t)$ be the probability that the site at origin is visited by
a walker for the first time at time $t$, whose initial position was 
$x$ at $t=0$. Within the MF approximation, any walker would survive
with probability $n(t)$, and the probability that it will make a first
visit to origin at time $t$ is given by 
$q(x,t)=\sqrt{\frac{2}{\pi}}\frac{x}{t^{3/2}}e^{-\frac{x^2}{2t}}$\cite{FELLER1}.
It follows that $R(x,t)=n(t)q(x,t)$.
We now integrate $R(x,t)$ over all initial positions $x$ and multiply
by the probability that the origin is persistent at time $t$, which is simply $P(t)$. 
So we find

\begin{equation}
\frac{\partial P(t)}{\partial t}=-P(t)n(t)K_{1}(t)~~~~~~~~d=1
\label{eq:EQ3}
\end{equation}

where $K_{1}(t)=\int_{-\infty}^{\infty}q(x,t)dx$ is the Smoluchowski constant\cite{SMOL} in $d=1$.
After substituting for $q(x,t)$ and $n(t)$, we find $\frac{\partial P(t)}{\partial t}=-\frac{\theta^{*}}{t}P(t)$,
so that $P(t)\sim t^{-\theta^{*}(q)}$, and $\theta^{*}(q)=\frac{\sqrt{2}}{\pi}\frac{q-1}{q}$ is the Persistence
exponent within the MF model\cite{STEVE2}. It is interesting to compare the mean field prediction for $\theta$ 
with the exact value
of the exponent. For $q=2$, $\theta^{*}(2)\simeq 0.225$, while the exact value from Eq.\ref{eq:EQ1} is $\frac{3}{8}$.
For the $q=\infty$ case, the MF model predicts $\theta^{*}(\infty)\simeq 0.45$, which is to be compared with the exact
value $\theta(\infty)=1$. Upon extending the comparison to the entire range of values of $q$, it is clear that
the mean field treatment consistently under-estimates the value of $\theta$. 

We now argue that $\theta^{*}(q)$ is, in fact, a lower bound for $\theta(q)$.
In the mean field approach discussed so far, it is assumed that the random walkers disappear from the 
lattice at random at such a rate so  that their average density falls as $n(t)$. The actual reaction-diffusion process is quite
different, because only walkers which come very close to another walker are likely to be removed.
Clearly in regions of space where walkers come close to each other, they are likely to visit
the same site again and again. This effect is much more within the mean field approach, where the walkers actually
pass through each other, possibly several times before disappearing. Thus, it is plausible that for the
same average density of walkers, a larger number of persistent sites will be visited in the actual reaction-diffusion 
model, compared to its mean field analogue.
Since this is true for all times, the average density of persistent sites in mean field theory
will be higher than the same in the actual Potts model dynamics. This would naturally imply that

\begin{equation}
\theta^{*}(q)\leq \theta(q)
\label{eq:EQ4}
\end{equation}

Interestingly, we show now that the Mean Field argument presented above yields the correct
value for the Persistence exponent for $A+A\to\emptyset$ model in $d=2$. This is not surprising,
since for this model, the upper critical dimension is $d_{c}=2$, and the mean field treatment becomes exact above
this dimension. It can be shown that in $d=2$, the probability that a random walker starting at an
an arbitrary point crosses a circle of radius $a$ around the origin for the first
time at $t$ is given by the expression\cite{SMOL2D}

\begin{equation}
K_{2}(t)\simeq \frac{4\pi D}{log(4Dt/a^{2})} ~~~~~~~t\gg a^{2}/D
\label{eq:EQ5}
\end{equation}

which is the analogue of Smoluchowski constant in $d=2$. 
The asymptotic particle density decay for $A+A\to\emptyset$ model in $d=2$ has the form 
$n(t)\simeq \frac{1}{8\pi}\frac{log(Dt)}{Dt}$\cite{BPLEE}. Upon extending the MF arguments presented
previously, we find that 

\begin{equation}
\frac{\partial P(t)}{\partial t}=-P(t)n(t)K_{2}(t)~~~~~~~d=2
\end{equation}

After substituting for $n(t)$ and $K_{2}(t)$, and taking the limit $a\to 0$, we find that 
$P(t)\sim t^{-\frac{1}{2}}$ so that $\theta^{*}=1/2$ in $d=2$. This result is exact, 
as has  been shown by a rigorous field-theoretic calculation\cite{CARDY}.

We thus observe that while the mean field approach, in general, gives only a lower bound for the 
persistence exponent, it correctly identifies the essential features that brings about this power-law
decay, ie., the diffusive motion of interfacial points and the $\frac{1}{\sqrt{t}}$ decay in
their over-all density. In the following sections, we use a slightly modified version of this treatment
to study the spatial distribution of persistent sites in $q$-state Potts model.

\section{The Modified Mean Field Approximation (MMFA)}

Our purpose is to use the mean field approach to study the spatial distribution of persistent spins
in the $q$-state Potts model, and in particular, to understand the transition from fractal to
homogeneous distribution as $\theta$ crosses $\frac{1}{2}$. However, it may be noted
that in the mean field approximation to the dynamics of the Potts model, the largest value of $\theta$
(attained at $q=\infty$) is $\frac{2}{\sqrt{\pi}}\simeq 0.45$ which is less than the transition
value $\frac{1}{2}$. This problem is circumvented by defining an artificial model where we define the 
diffusing particles as non-interacting random walkers, who can pass through each other. The model
also allows for multiple occupancy of lattice sites. The dynamics consists of random walkers being
picked at random and taken out of the lattice at a time-dependent rate, which is tuned to produce 
power-law decay $P(t)\sim t^{-{\theta}^{\prime}}$ with any arbitrary ${\theta}^{\prime}$. From
the arguments presented in the previous section, it is obvious that this would be the case 
if the average density were to 
decay as $n(t)\sim \sqrt{\frac{\pi}{2}}\frac{{\theta}^{\prime}}{\sqrt{t}}$ asymptotically. 
By construction, this model is devoid of spatial correlations among reacting particles 
(ie., it is still mean field) but ${\theta}^{\prime}$ is now arbitrary. If we now choose 
${\theta}^{\prime}=\theta(q)$, this model is an approximation to the $q$-state Potts model, with
the simplifying feature that the spatial correlation between interfacial points is now absent.
We shall refer to this model as the Modified Mean Field Approximation (MMFA) for the original
Potts model.

\subsection{Pair correlation for persistent spins within the MMFA}

In this section, we compute the equal time pair correlation function for persistent
spins in $q$-state Potts model under the MMFA.  We define $P_{2}(r,t)$ as the probability that 
both the site at origin and the site at $r>0$ are persistent at time $t$. Our purpose is
to compute $P_{2}(r,t)$ for various $r$.

The generalization of Eq.\ref{eq:EQ3} to this case is

\begin{equation}
-\frac{\partial P_{2}(r,t)}{\partial t}=2P_{2}(r,t)\int_{-\infty}^{r}R_{r}(x,t)dx
\label{eq:EQ6}
\end{equation}

where $R_{r}(x,t)$ is the probability that a particle with initial position
$x (-\infty < x < r)$ will make a first visit to the origin at time $t$, {\it without
ever crossing $r$ in the interval [0:t]}.
The factor 2 in front takes into account the probability that either of the sites
could be reached by one of the diffusing particles. Unlike the first case,
$R_{r}(x,t)$ is now different for $x<0$ and $0\leq x<r$. For $x<0$, the constraint
of no crossing at $r$ is irrelevant for the computation of $R_{r}(x,t)$ , since to 
reach $r$, the particle would have to cross the origin first. So $R_{r}(x,t)=R(x,t)$
simply for $x<0$, and so 

\begin{equation}
\int_{-\infty}^{0}R_{r}(x,t)dx=\frac{\theta}{2t}
\label{eq:EQ7}
\end{equation}

For $x>0$, this is no longer true, and $R_{r}(x,t)$ needs to be computed
separately. The quantity that we need here is $q_{r}(x,t)$, the probability
that a diffusing particle whose position at $t=0$ is $x$, will reach the origin
for the first time at $t$, without ever crossing the point $r$ in between.
Then $R_{r}(x,t)=n(t)q_{r}(x,t)$.
To find $q_{r}(x,t)$, let us use the following standard method. Consider a random
walk starting from $0< x < r$ at $t=0$ with absorbing barriers at $0$ and $r$.
If the probability distribution of the position $z$ of the walker at time $t$ is
$u_{x}(z,t)$, then 

\begin{equation}
q_{r}(x,t)=\frac{\partial u_{x}(z,t)}{\partial z}|_{z=0}
\label{eq:EQ8}
\end{equation}

The expression for $u_{x}(z,t)$ is known exactly, and the asymptotic form at
large $t$\cite{FELLER1} is

\begin{equation}
u_{x}(z,t)=\frac{1}{\sqrt{2\pi t}}\sum_{k=-\infty}^{\infty}e^{-\frac{(z-x-2kr)^2}{2t}}
-e^{-\frac{(z+x-2kr)^2}{2t}}
\label{eq:EQ9}
\end{equation}

from which, we find

\begin{eqnarray}
q_{r}(x,t)=\frac{1}{\sqrt{2\pi t}}\sum_{k=-\infty}^{\infty}
\frac{x+2kr}{t}e^{-\frac{(z-x-2kr)^2}{2t}}-\nonumber \\
\frac{x-2kr}{t}e^{-\frac{(z+x-2kr)^2}{2t}}
\label{eq:EQ10}
\end{eqnarray}

We note that for $r\gg t^{\frac{1}{2}}$, the $k=0$ mode is the dominant term in
the sum, and this gives $q_{r}(x,t)\approx \sqrt{\frac{2}{\pi t^3}}xe^{-\frac{x^2}{2t}}$+
smaller terms that vanish as $\frac{r}{t^{1/2}}\to\infty$. Clearly in this limit, we
recover the $r=\infty$ term, as we should. It then follows that

\begin{equation}
\int_{0}^{r}R_{r}(x,t)dx=\frac{\theta}{2t}G\left(\frac{r}{\sqrt{t}}\right)
\label{eq:EQ11}
\end{equation}

where

\begin{equation}
G(x)=1-\eta+\sum_{k=1}^{\infty}2\eta^{4k^2}-\eta^{(1+2k)^2}-\eta^{(1-2k)^2}
\label{eq:EQ12}
\end{equation}

with $\eta=e^{-\frac{x^2}{2}}$. After substitution of Eq.\ref{eq:EQ7} and Eq.\ref{eq:EQ11} 
 in Eq. \ref{eq:EQ6}, we find 

\begin{figure}
\epsfxsize=2.3in
\epsfbox{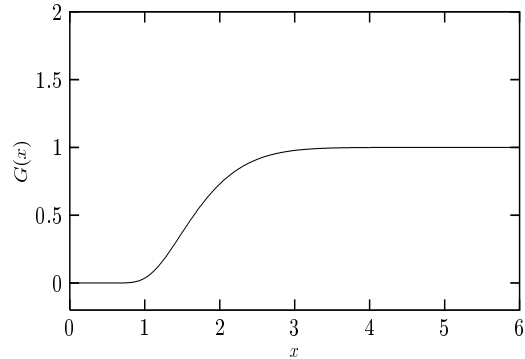}
\caption{Plot of $G(x)$ vs $x$}
\end{figure}

\begin{equation}
\frac{\partial P_{2}(r,t)}{\partial t}=-2P_{2}(r,t)\frac{\theta}{2t}\left[1+G(\frac{r}{\sqrt{t}})\right]
\label{eq:EQ13}
\end{equation}

which admits a scaling solution of the form

\begin{equation}
P_{2}(r,t)=P(t)^{2}f(\frac{r}{\sqrt{t}})
\label{eq:EQ14}
\end{equation}

and the scaling function $f(x)$ is given by the following expression.

\begin{equation}
\frac{x}{2}\frac{\partial f}{\partial x}=-\theta f(x)\left[1-G(x)\right]
\label{eq:EQ15}
\end{equation}

Let us now consider the limiting behaviour of the scaling function for $x\ll 1$ and
$x\gg 1$. In the first case, it is clear from Fig.1 that $G(x)\approx 0$, and so
$\frac{x}{2}\frac{\partial f}{\partial x}=-\theta f(x)$, which implies that
$f(x)\sim x^{-2\theta}$ as $x\to 0$. In the opposite extreme $G(x)\to 1$ as $x\to\infty$,
and so $\frac{x}{2}\frac{\partial f}{\partial x}\approx 0$, which means that
$f(x)$ approaches a constant value in this limit. It is further clear that, from the definition of the scaling
function as given by the expression Eq.\ref{eq:EQ14}, this constant is unity, since
we expect $P_{2}(r,t)\to P(t)^{2}$ as $r\to\infty$. 
For convenience of later calculations, we approximate the scaling function as 

\begin{eqnarray}
f(x)= a^{2\theta}x^{-2\theta}~~~~~~:x\leq a\nonumber \\
f(x)= 1~~~~~~~~~~:x> a
\label{eq:EQ16}
\end{eqnarray}

where $a$ is a number, of order unity.

We see that under the MMFA, the pair correlation function has a scaling form which is same for 
all values of $\theta$, with power-law decay $P_{2}(r,t)\sim P(t)r^{-2\theta}$ for short
distances $r\ll t^{\frac{1}{2}}$.
As is well-known, power law decay of pair correlation function points to the underlying
scale invariance of the spatial distribution of the persistent spins. This is characteristic
of a fractal distribution under some circumstances. To see this, let us first define
$C(r,t)=P(t)^{-1}P_{2}(r,t)$, which is the probability of finding a persistent spin at
a distance $r$ from another persistent spin. Now, the integral $M(R,t)=\int_{1}^{R}C(r,t)dr$
is the total number of persistent spins within a radius $R$ of a persistent spin. Clearly,
from the scaling form described above, $M(R,t)\sim R^{d_{f}}$ for $R\ll t^{\frac{1}{2}}$,
where $d_{f}=$max$(1-2\theta,0)$. For $R\gg t^{\frac{1}{2}}$, we find that $M(R,t)\simeq RP(t)$,
which is simply a homogeneous distribution. Thus, if we look at length scales
$R\ll t^{\frac{1}{2}}$, there is a fractal structure when $\theta < \frac{1}{2}$. However, 
when $\theta\geq \frac{1}{2}$, this scale-invariant structure is replaced by a few isolated sites, 
whose number does not grow with the length scale of observation. 

Clearly, the spatial distribution of persistent
spins undergoes a transition as $\theta$ crosses $\frac{1}{2}$. Indeed, if we consider time
scales beyond equilibration time $t\gg L^{2}$, for $\theta < \frac{1}{2}$ the total
number of persistent spins left in the system scales as $L^{1-2\theta}$, whereas for 
$\theta\geq \frac{1}{2}$, there are only a finite number of persistent spins left.
This important difference is not adequately reflected in the pair correlation function, which
has the same scaling form for all values of $\theta$. In the next section, we study another
quantity to characterize the spatial distribution which undergoes a rather significant
change in its scaling properties across the transition. This quantity is the empty interval
distribution, which is one of the standard tools in the study of one-dimensional reaction-diffusion
processes.

\subsection{The Empty Interval Distribution}

An empty interval, in our convention, is the separation between two consecutive persistent
sites. The empty interval distribution (EID) $n(k,t)$ is defined as the number of such intervals
of length $k$ at time $t$. For convenience, we also divide this quantity with the system
size $N$ so that $n(k,t)$ satisfies the following normalization conditions.

\begin{equation}
\int_{1}^{\infty}n(k,t)dk=P(t)\hspace{0.3cm};\int_{1}^{\infty}kn(k,t)dk=1
\label{eq:EQ17}
\end{equation}

Computing the EID directly, even under the mean field approximation, is non-trivial.
Instead we shall compute it from the pair correlation function using the Independent
Interval Approximation (IIA), where the lengths of successive empty intervals are
considered as independent random variables. The IIA has been a valuable tool in the study
of one-dimensional problems, and has been successfully applied to study spatial distribution
of persistent spins in $A+A\to\emptyset$ model.
Under the IIA, the relation between $n(k,t)$ and $P_{2}(r,t)$ is 

\begin{equation}
P_{2}(r,t)=n(r,t)+P(t)^{-1}\int_{1}^{r}dx n(x,t)P_{2}(r-x,t)
\label{eq:EQ18}
\end{equation}

It is convenient to express this relation in terms of the Laplace
transforms ${\tilde C}(p,t)=\int_{1}^{\infty}C(r,t)e^{-pr}dr$ and ${\tilde n}(p,t)=
\int_{1}^{\infty}n(s,t)e^{-ps}ds$, where $C(r,t)$ was defined in the previous
section. Under these transformations, Eq.\ref{eq:EQ18} maybe expressed in the form

\begin{equation}
{\tilde n}(p,t)=\frac{P(t){\tilde C}(p,t)}{1+{\tilde C}(p,t)}
\label{eq:EQ19}
\end{equation}

From the scaling form for $P_{2}(r,t)$ given by Eq.\ref{eq:EQ14} , we find that

\begin{equation}
{\tilde C}(p,t)=P(t)\sqrt{t}I(q,t)
\label{eq:EQ20}
\end{equation}

where $q=p\sqrt{t}$, and $I(q,t)=\int_{t^{-\frac{1}{2}}}^{\infty}f(x)e^{-qx}dx$
The lower limit is put as $t^{-\frac{1}{2}}$ instead of zero to
take care of possible small argument divergence in the scaling function.

Let us first consider the case where $\theta< \frac{1}{2}$:
In this case the scaling function $f(x)$ is integrable, so we put the lower limit
in the previous equation as zero. Using Eq.\ref{eq:EQ16}, we find that

\begin{equation}
I(q,t)=a^{2\theta}q^{2\theta-1}\gamma(1-2\theta,qa)+\frac{1}{q}e^{-qa}~~~~~; {\theta} < \frac{1}{2}
\label{eq:EQ21}
\end{equation}

where $\gamma(\alpha,x)=\int_{0}^{x}e^{-t}t^{\alpha-1}dt$ is the incomplete Gamma function.
After substituting in Eq.\ref{eq:EQ19} and Eq. \ref{eq:EQ20}, and taking the 
$t\to\infty$ limit (keeping $q$ fixed), we find




\begin{equation}
{\tilde n}(p,t)=t^{-\frac{1}{2}}\left[P(t)t^{\frac{1}{2}}-
\frac{q}{a^{2\theta}q^{2\theta}\gamma(1-2\theta,qa)+e^{-qa}}\right]
\label{eq:EQ22}
\end{equation}

It follows that 

\begin{equation}
n(k,t)=t^{-1}h\left(\frac{k}{t^{\frac{1}{2}}}\right)
\label{eq:EQ23}
\end{equation}

so that 

\begin{equation}
{\tilde n}(p,t)=t^{-\frac{1}{2}}[t^{\frac{1}{2}}P(t)-h_{1}(q)]
\label{eq:EQ24}
\end{equation}

which has the same form as Eq.\ref{eq:EQ22}, and 
$h_{1}(q)=\int_{0}^{\infty}h(x)[1-e^{-qx}]dx$. After integrating by parts, we find

\begin{equation}
h_{1}(q)=-G(\infty)+q\left[\lim_{x\to 0}xG(x)+{\tilde G}(q)\right]
\label{eq:EQ25}
\end{equation}

where $G(x)=\int_{x}^{\infty}h(y)dy$ and ${\tilde G}(q)=\int_{0}^{\infty}G(x)e^{-qx}dx$.
We assume that $G(x)$ is integrable, so that $\lim_{x\to 0}xG(x)=0$ and $G(\infty)=0$.
Finally we have 

\begin{equation}
{\tilde G}(q)=\frac{1}{a^{2\theta}q^{2\theta}\gamma(1-2\theta,qa)+e^{-qa}}
\label{eq:EQ26}
\end{equation}

Now we may try to deduce the behaviour of the function $G(x)$ at large and small
arguments from its Laplace transform, given by the previous equation. To find
the behaviour of $G(x)$ near $x=0$, we use the standard formula\cite{LAPLACE}

\begin{equation}
lim_{t\to 0}t^{-\rho}g(t)=lim_{s\to\infty}\frac{s^{\rho+1}{\tilde g(s)}}{\rho!}~~~~;\rho > -1
\label{eq:EQ27}
\end{equation}

where ${\tilde g(s)}$ is the L.T of $g(t)$. Now,
for large $q$, $\gamma(1-2\theta,qa)\simeq \Gamma(1-2\theta)$, so that 
${\tilde G}(q)=\frac{a^{-2\theta}q^{-2\theta}}{\Gamma(1-2\theta)}$ as $q\to\infty$. 
It follows that $lim_{x\to 0}x^{1-2\theta}G(x)=\frac{a^{-2\theta}}{(2\theta-1)!\Gamma(1-2\theta)}$
from which we find

\begin{equation}
G(x)\sim \frac{a^{-2\theta}x^{2\theta-1}}{(2\theta-1)!\Gamma(1-2\theta)}~~~~x\to 0
\label{eq:EQ28}
\end{equation}

and, after using the relation $h(x)=-\frac{\partial G}{\partial x}$, 

\begin{equation}
h(x)\sim x^{-2(1-\theta)}~~~~; x\to 0
\label{eq:EQ29}
\end{equation}

The behaviour of $G(x)$ at large $x$, one has to look for the singularities of
${\tilde G}(q)$ in Eq. \ref{eq:EQ26}. If ${\tilde G}(q)$ has a 
singularity of the form ${\tilde G}(q)\sim (q-q^{*})^{-\nu}$, then, upon inversion
of the L.T, it follows that $G(x)\sim x^{\nu-1}e^{q^{*}x}$ as $x\to\infty$\cite{LAPLACE}.
In order to find the singularity, we plotted the denominator of 
Eq. \ref{eq:EQ26} against its argument (Fig. 2). We find that the function
crosses zero at one point in the negative $q$ axis. By careful numerical analysis using
bisection method, we
have determined this crossing point to be at $q^{*}a=-\lambda$, where the numerical 
constant $\lambda\simeq 0.32$ for $\theta=\frac{3}{8}$ and $\lambda\simeq 0.85$ for
$\theta=\frac{1}{2}$. This implies that the leading term in the decay of $G(x)$ at large $x$ is
exponential, ie., $G(x)\sim exp(-\frac{\lambda}{a}x)$ as $x\to \infty$, with a possible 
power-law prefactor. Consequently, the scaling function $h(x)$ also has similar exponential
decay at large $x$. 

\begin{equation}
h(x)\sim e^{-\frac{\lambda}{a}x}~~~~~~; x\gg 1
\label{eq:EQ30}
\end{equation}

\begin{figure}
\epsfxsize=2.3in
\epsfbox{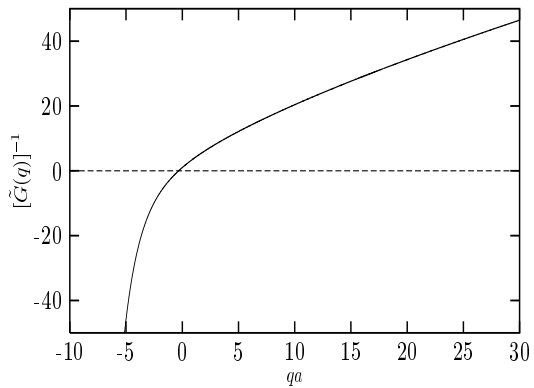}
\caption{The figure shows the inverse of ${\tilde G}(q)$ plotted against $qa$.}
\end{figure}

We also determine the characteristic length scales of the distribution using
the scaling form for $n(k,t)$. These may be defined as the ratios of moments
of the distribution.

\begin{equation}
L_{m}(t)=I_{m}(t)/I_{m-1}(t)~~~~~; m=0,1,2,......
\label{eq:EQ31}
\end{equation}

where $I_{m}(t)=\sum_{k}n(k,t)k^{m}$ are the moments of $n(k,t)$.
Clearly, $L_{1}(t)=P(t)^{-1}\sim t^{\theta}$ by definition, whereas all higher
order length scales 

\begin{equation}
L_{j}(t)\sim t^{\frac{1}{2}}~~~~~; j >1
\label{eq:EQ32}
\end{equation}

which follows from the dynamic scaling form given by Eq.\ref{eq:EQ23}, \ref{eq:EQ29} and \ref{eq:EQ30} 
for $n(k,t)$.

We now continue our study of empty interval distribution for the case where $\theta\geq \frac{1}{2}$.
Although the MMFA allows us to study arbitrarily large values of $\theta$, we restrict ourselves
to the regime $\theta < 1$, since our basic aim is to study the persistence in $q$-state Potts model
where $\theta(q)\leq 1$. Furthermore, for $q=\infty$ where $\theta=1$, $n(k,t)$ can be
found exactly\cite{STEVE1} and is known to be a pure exponential.

For $\frac{1}{2}< \theta < 1$, the scaling function $f(x)$ has a non-integrable $x^{-2\theta}$ 
singularity near $x=0$ (We do not study explicitly the logarithmic singularity occurring
for $\theta = \frac{1}{2}$). We integrate by parts and find

\begin{eqnarray}
I(q,t)=\frac{a^{2\theta}}{2\theta-1}[t^{\theta-\frac{1}{2}}e^{-qt^{-\frac{1}{2}}}-a^{1-2
\theta}e^{-qa}-\nonumber \\
q^{2\theta-1}\gamma(2-2\theta,qa)]+\frac{1}{q}e^{-qa}~~~~~; {\theta}\geq \frac{1}{2}
\label{eq:EQ33}
\end{eqnarray}

Let us now define $\lambda=\frac{p}{P(t)}=qt^{-\frac{1}{2}}P(t)^{-1}$. For $t\to\infty$ and
finite $\lambda$, we have

\begin{equation}
I(q,t)=\frac{a^{2\theta}}{2\theta-1} t^{\theta-\frac{1}{2}}+\frac{1}{\lambda t^{\frac{1}{2}}P(t)}
\label{eq:EQ34}
\end{equation}

After substitution in Eq.\ref{eq:EQ20} we find

\begin{equation}
{\tilde C}(p,t)=P(t)\left[\frac{a^{2\theta}}{2\theta-1}t^{\theta}+\frac{1}{\lambda P(t)}\right]
\label{eq:EQ35}
\end{equation}

We define the constant $\beta=\frac{t^{\theta}P(t)a^{2\theta}}{2\theta-1}$, in terms of which
${\tilde C}(p,t)=\beta+\frac{1}{\lambda}$. Now we substitute in Eq.\ref{eq:EQ19} and find

\begin{equation}
{\tilde n}(p,t)=P(t)\frac{\beta}{1+\beta}\left[\frac{\lambda+\beta^{-1}}{\lambda+(1+\beta)^{-1}}\right]
\label{eq:EQ36}
\end{equation}

Upon inversion of the L.T, we find that 

\begin{equation}
n(k,t)=P(t)^{2}\phi[kP(t)]
\label{eq:EQ36A}
\end{equation}

where 

\begin{equation}
\phi(x)=\frac{\beta}{1+\beta}\left[\delta(x)+\frac{1}{\beta(1+\beta)}e^{-x(1+\beta)^{-1}}\right]
\label{eq:EQ36B}
\end{equation}

The scaling function has a rather unnatural $\delta$-function singularity at the origin. 
However, a more careful analysis show that for any finite (but still large) time $t$, the
divergence at origin is only power-law, but with a different exponent than the
previous case ($\theta < \frac{1}{2}$). We start with the expression given by
Eq.\ref{eq:EQ20} and Eq.\ref{eq:EQ33} for ${\tilde C}(p,t)$.
After keeping the leading finite $t$ correction, we find that

\begin{equation}
{\tilde C}(p,t)=\beta+\frac{1}{\lambda}-
{\beta}t^{-\theta(2\theta-1)}\lambda^{2\theta-1}
\label{eq:EQ37}
\end{equation}

For purposes that will be clear later, let us define $m(k,t)=kn(k,t)$ so that
$\sum_{k}m(k,t)=1$. We also define the associated Laplace transform ${\tilde m}(p,t)$.
The Laplace transforms are related through

\begin{equation}
{\tilde m}(p,t)=-\frac{\partial {\tilde n}(p,t)}{\partial p}
\label{eq:EQ38}
\end{equation}

Using the expression Eq.\ref{eq:EQ19} for ${\tilde n}(p,t)$, we find that

\begin{equation}
{\tilde m}(p,t)=-\frac{P(t){\tilde C}^{\prime}(p,t)}{[1+{\tilde C}(p,t)]^{2}}
\label{eq:EQ39}
\end{equation}

where ${\tilde C}^{\prime}(p,t)=\frac{\partial {\tilde C}(p,t)}{\partial p}$, and is given by
the expression

\begin{equation}
{\tilde C}^{\prime}(p,t)=-\frac{1}{P(t)}
\left[\frac{1}{\lambda^{2}}+(2\theta-1)\beta t^{-\theta(2\theta-1)}\lambda^{2(\theta-1)}\right]
\label{eq:EQ40}
\end{equation}

After substitution in Eq.\ref{eq:EQ38} and taking the limit $t\to\infty$, we find

\begin{equation}
{\tilde m}(p,t)=\frac{1+(2\theta-1)\beta t^{-\theta(2\theta-1)}\lambda^{2\theta}}
{[1+\lambda(1+\beta)]^{2}}
\label{eq:EQ41}
\end{equation}

which gives the scaling forms 

\begin{equation}
m(k,t)=P(t)\psi(t,kP(t))~~~;~~~n(k,t)=P(t)^{2}\Phi(t,kP(t))
\label{eq:EQ42}
\end{equation}

where 

\begin{equation}
x\Phi(t,x)=\psi(t,x)
\label{eq:EQ43}
\end{equation}

by definition. The Laplace transform of the scaling function $\psi(t,x)$ is

\begin{equation}
{\tilde \psi}(t,\lambda)=\frac{1+(2\theta-1)\beta t^{-\theta(2\theta-1)}\lambda^{2\theta}}
{[1+\lambda(1+\beta)]^{2}}
\label{eq:EQ44}
\end{equation}

We notice that if the finite $t$ correction term is not included, 
$lim_{\lambda\to\infty}\lambda^{2}{\tilde \psi}(t,\lambda)$ is finite, and in that case, the
small argument divergence of $\psi(t,x)$ will be sharper than any power-law. This is what
is reflected in the appearance of the $\delta$-function in Eq.\ref{eq:EQ38}. 
However, when this term is
included, the multiplying factor has to be $\lambda^{2-2\theta}$ in order to make the resulting
expression finite as $\lambda\to\infty$. This implies that the small $x$ divergence for $\psi(t,x)$
has the power-law form $\psi(t,x)\sim t^{-\theta(2\theta-1)}x^{1-2\theta}$ for small $x$. From
Eq. \ref{eq:EQ45}, we find a similar power-law divergence in $\Phi(t,x)$ also.

\begin{equation}
\Phi(t,x)\sim t^{-\theta(2\theta-1)}x^{-2\theta}~~~~~; x\ll 1
\label{eq:EQ45}
\end{equation}

In the large $x$ limit, $\Phi(t,x)$ becomes time independent, and decays exponentially
with $x$ as in Eq. \ref{eq:EQ36B}, ie., 

\begin{equation}
\Phi(t,x)\simeq \frac{1}{(1+\beta)^2}e^{-\frac{x}{1+\beta}}~~~~; x\gg 1
\label{eq:EQ46}
\end{equation}

The characteristic length scales are easy to compute. From the scaling form, it
follows that all the characteristic lengths have identical asymptotic scaling
behaviour.

\begin{equation}
L_{j}(t)\sim t^{\theta}~~~~~; j=1,2,...
\label{eq:EQ47}
\end{equation}

The difference in the asymptotic scaling behaviour of the characteristic length
scale as $\theta$ crosses $\frac{1}{2}$ may be seen as a competition between two length scales, 
the diffusive length scale ${\cal L}_{D}(t)\sim \sqrt{Dt}$ which gives the 
mean separation between two random walkers, and the persistence scale 
${\cal L}_{p}(t)=P(t)^{-1}\sim t^{\theta}$ which is the mean separation between two
persistent spins. The characteristic length scale is dominated by
the larger of the two, ie., we may write 

\begin{equation}
{\cal L}(t)\sim t^{\phi}~~~~; \phi=$max$(\frac{1}{2},\theta)
\label{eq:EQ48}
\end{equation}

where ${\cal L}(t)$ is defined through the dynamical scaling form for $n(k,t)$.

\begin{equation}
n(k,t)={\cal L}(t)^{-2}g(t,\frac{k}{{\cal L}(t)})
\label{eq:EQ49}
\end{equation}

The scaling function $g(t,x)=h(x)$ when $\theta < \frac{1}{2}$ and $g(t,x)=\Phi(t,x)$ when
$\theta \geq \frac{1}{2}$. In general, the small argument behaviour of $g(t,x)$ has the
power-law form

\begin{equation}
g(t,x)\sim t^{-\psi}x^{-\tau}~~~~; x\to 0
\label{eq:EQ50}
\end{equation}

where the exponents $\psi$ and $\tau$ are given by

\begin{equation}
\psi=\theta(2\theta-1)H(\theta-\frac{1}{2})~~;~~\tau=$max$[2\theta,2(1-\theta)],
\label{eq:EQ51}
\end{equation}

and $H(x)$ is the Heaviside step function. For large $x$, the scaling function is time-independent 
and decays exponentially with $x$.
We also note from the scaling form that over small distances $k\ll t^{\phi}$, 

\begin{equation}
n(k,t)\sim P(t)k^{-\tau}~~~~;k\ll t^{\phi}
\label{eq:EQ52}
\end{equation}

where $\tau$ is given by Eq.\ref{eq:EQ51}. 

\section{NUMERICAL RESULTS}

We studied the quantities $P_{2}(r,t)$ and $n(k,t)$ numerically by simulating the kinetics
of $q$-state Potts model with random initial conditions. The time evolution of spin
configurations via Glauber dynamics is implemented using the mapping of this dynamics
to a reaction-diffusion problem, as mentioned in the introduction. In this procedure,
a set of diffusing particles $A$ are initially distributed at random on the lattice
with a certain average initial density $n_{0}$ (which we fix as $\frac{1}{2}$). When two
diffusing particles meet, they annihilate each other or coagulate with probability
$\frac{1}{q-1}$ and $\frac{q-2}{q-1}$ respectively. We count one MC step in the simulation
after the position of every particle in the lattice has been updated once. Persistent
spins(sites) at any time $t$ are those sites which have not been touched by a random
walker till that time. All the simulations were done on a lattice with $2^{17}$ sites, and 
the results were averaged over 100 different starting configurations. In order to check the
different dynamic scaling behaviour for $\theta <\frac{1}{2}$ and $\theta\geq \frac{1}{2}$,
we did our simulations for three different values of $q$-2,5 and 10. For later reference, we
note that from Eq.\ref{eq:EQ1}, the corresponding values of the persistence exponent are
$\theta(2)=3/8=0.375, \theta(5)\approx 0.6928$ and $\theta(10)\approx 0.8310$. 
In Figs. 3-5, and later in Figs. 7-10, we have employed
logarithmic binning of the data in intervals of size $1.5^{n}$ ($n=1,2,....$). 
since the statistical noise was considerable. However, for all exponent measurements, we have
used only the bare(not binned) data.

In Fig. 3-5, we have plotted the scaling function $f(x)$ for the pair correlation function
$P_{2}(r,t)$ against  the scaling variable $r/\sqrt{t}$ for three $q$-values, $q=2,5$ and 10. 
We find excellent scaling collapse for all three values of $q$, which is in agreement with
the dynamic scaling picture provided by the MMFA in Eq.\ref{eq:EQ14}. 
In the figures, we find power-law decay of $f(x)$ for small $x$, with a sharp cross-over to 
the flat long-distance behaviour, which is also in agreement with the assumption we made
in Eq.\ref{eq:EQ16}. We also note that the constant $a$ introduced in Eq.\ref{eq:EQ16} 
is in fact very close to 1. 

\begin{figure}
\epsfxsize=1.8in
\epsfbox{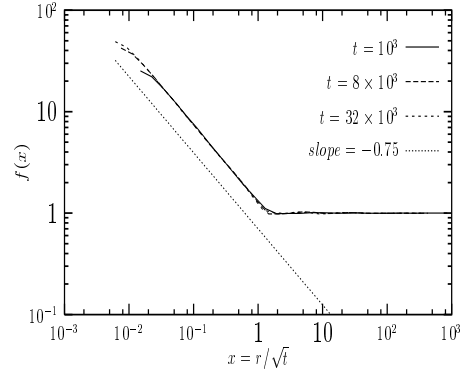}
\caption{The scaled pair correlation $f(x)=P(t)^{-2}P_{2}(r,t)$ is plotted against
the dimensionless scaling variable $x=r/\sqrt{t}$ for $q=2$ on a logarithmic scale. The straight line
is a guide to eye, and has slope $2\theta(2)=0.75$ , which is the MMFA prediction. The time $t$ is
measured in MC steps and distance $r$ is measured in units of lattice spacing.}
\end{figure}

\begin{figure}
\epsfxsize=1.8in
\epsfbox{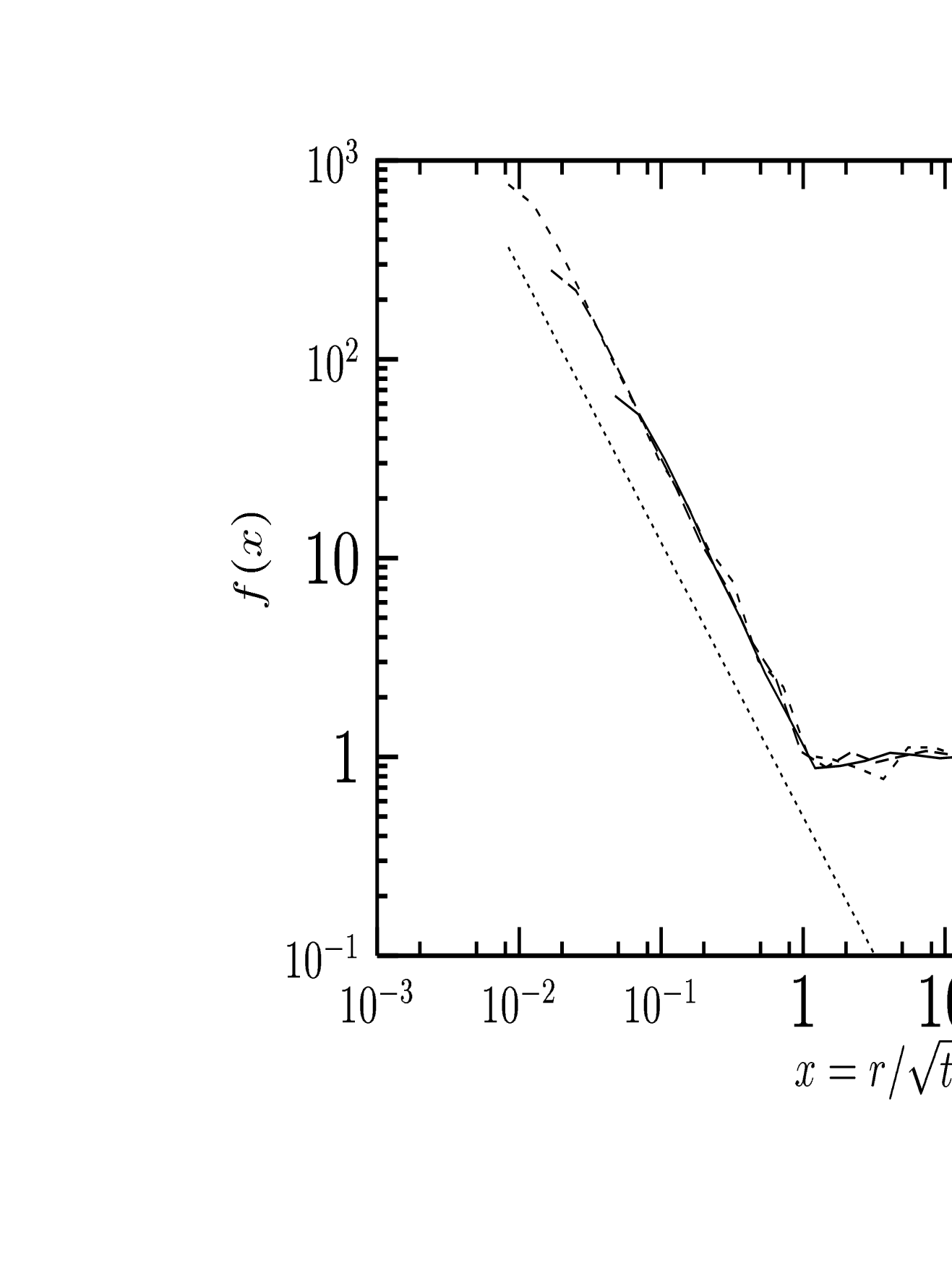}
\caption{Same as Fig.3, for $q=5$. The slope of the straight line is $2\theta(5)\simeq 1.38$,
which is the MMFA prediction.}
\end{figure}

\begin{figure}
\epsfxsize=1.8in
\epsfbox{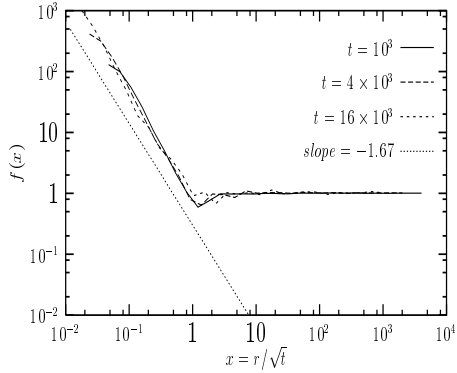}
\caption{Same as Fig.3, for $q=10$.The slope of the straight line is $2\theta(10)\simeq 1.67$,
which is the MMFA prediction.}
\end{figure}

In Fig. 6, we plot the characteristic length scale $L_{2}(t)$ against time $t$ for $q=2,5$
and 10 and measure the dynamical exponent $\phi$. 
The observed slopes of the lines are systematically higher than the theoretical
prediction in Eq.\ref{eq:EQ50} by around 0.05, while the statistical error in all the
three cases was only $\sim 10^{-4}$ or smaller. The observed deviation could be possibly
due to the fact that the asymptotic behaviour is not fully reflected over the time scales
which we used. The number of persistent spins left in the system falls rapidly with time for
high values of $q$, and so we were forced to restrict ourselves to times $t\leq 32000$. In fact,
even for $q=2$ case, previous simulations over longer time scales have shown the presence
of an additive power-law correction to the asymptotic scaling behaviour\cite{MANOJ2}.

\begin{figure}
\epsfxsize=1.8in
\epsfbox{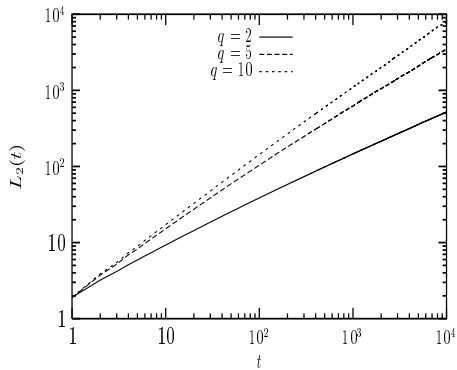}
\caption{The figure shows the characteristic length scale $L_{2}(t)$ (measured in units
of lattice spacing, definition
in text) plotted against time $t$ (measured as number of MC steps) on a logarithmic scale 
for three Potts values $q=2,5$ and 10. The measured slopes of
the lines are respectively 0.5507, 0.7391 and 0.8672. The corresponding 
theoretical predictions are, to the same accuracy, 0.5000, 0.6928 and 0.8310. }
\end{figure}

\begin{figure}
\epsfxsize=1.8in
\epsfbox{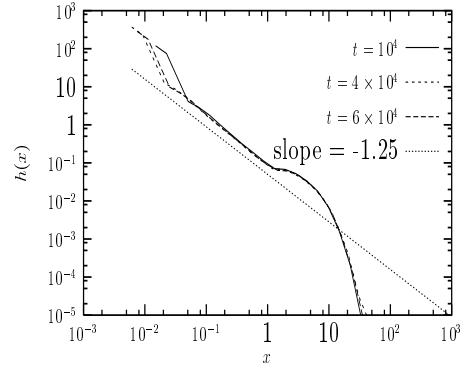}
\caption{The scaled EID $h(x)=tn(k,t)$ is plotted against 
the dimensionless scaled separation $x=k/\sqrt{t}$ for $q=2$ Potts model. The excellent scaling collapse
validates the scaling form given in Eq. \ref{eq:EQ23}. The straight line has slope 
$\tau=2(1-\theta(2))=5/4$, which is the MMFA-IIA prediction.The time $t$ is
measured in MC steps and distance $k$ is measured in units of lattice spacing.}
\end{figure}

In Fig. 7-9, we check the dynamic scaling form Eq.\ref{eq:EQ51} for $n(k,t)$ against the scaling
variable $x=k/t^{\phi}$ for three values of $q$--2,5 and 10. We find that for
$q=2$, excellent scaling collapse is obtained with $\phi=\frac{1}{2}$ (Fig. 7). For
small $x$, we find power-law decay of the scaling function, which crosses
over to fast exponential decay at large $x$. For higher values of $q$ 
(where $\theta(q) > \frac{1}{2}$), we find that with
$\phi=\theta$, we find very good scaling collapse for $x\gg 1$. But for $x\ll 1$,
we find systematic deviation from scaling collapse, which was also observed earlier 
by Bray and O'Donoghue\cite{STEVE1}. 
This observation supports the theoretical prediction based on MMFA, and shows that in this regime, the scaling 
function has explicit time dependence. To show this more clearly, and to verify
the predicted time-dependence, we plotted the quantity $n(k,t)/P(t)$ against $k$
for three widely spaced values of $t$ for all three $q$ values studied in Fig.9.
We see that in all three cases, a simple power-law decay with $k$ is observed for $k\ll t^{\phi}$, 
thus validating Eq.\ref{eq:EQ52}. The measurement of the exponent $\tau$ gives
values in reasonable agreement with theoretical prediction, although for $q=10$, the
statistical error is significant.

\begin{figure}
\epsfxsize=1.8in
\epsfbox{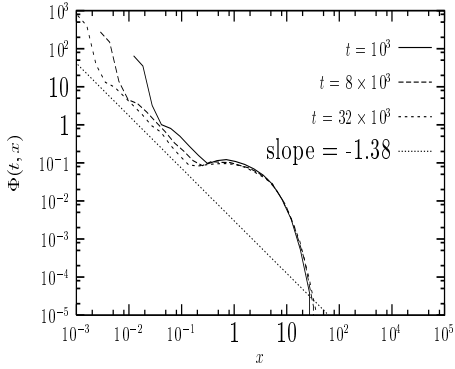}
\caption{The scaled EID $\Phi(t,x)=tn(k,t)$ is plotted against 
the scaled separation $x=kP(t)$ on a logarithmic scale for $q=5$ Potts model. We see that the 
scaling function is explicitly time-dependent for small $k$, but is time-independent for large $k$. 
The straight line in the figure gives the theoretical prediction $\tau=2\theta(5)\simeq 1.38$
for power-law decay at small $x$ (see discussion in text). The time $t$ is
measured in MC steps and distance $k$ is measured in units of lattice spacing.}
\end{figure}

\begin{figure}
\epsfxsize=1.8in
\epsfbox{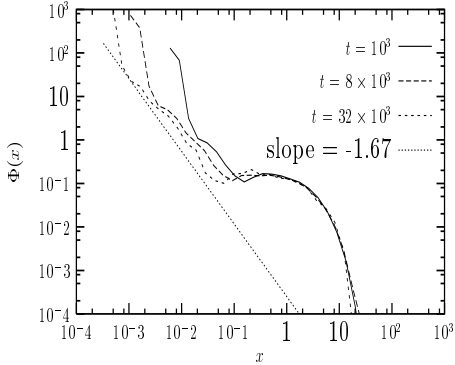}
\caption{Same as Fig. 6, for $q=10$. The straight line in the figure gives the 
theoretical prediction $\tau=2\theta(10)\simeq 1.67$
for power-law decay at small $x$ (see discussion in text). The time $t$ is
measured in MC steps and distance $k$ is measured in units of lattice spacing.}
\end{figure}

\begin{figure}
\epsfxsize=1.8in
\epsfbox{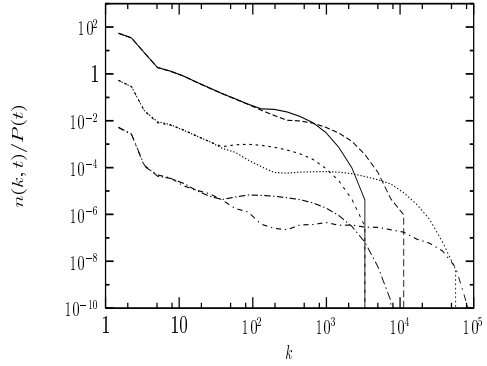}
\caption{In the figure, $n(k,t)/P(t)$ is plotted against $k$ for two widely separated values of
$t$ for each value of $q=2,5$ and 10 
 (top to bottom). In all the cases, the function is independent of $t$ for $k\ll t^{\phi}$,
and shows the power law decay $\sim k^{-\tau}$. We measure $\tau\simeq 1.32\pm 0.03$, $1.41\pm 0.04$ and
$1.61\pm 0.22$ for $q=2$, $q=5$ and $q=10$ respectively. 
The corresponding MMFA-IIA predictions are, to the same accuracy, 1.25, 1.38 and
1.66.}
\end{figure}

\section{Conclusions}
In this paper, we have studied the spatial aspects of persistence in one dimensional $q$-state
Potts model using a mean field approximation. We have computed the pair correlation function 
for persistent spins under this
approximation, and used it to compute the empty interval distribution under the independent interval
approximation. We find dynamical scaling behaviour in 
both these quantities. The time dependence of the characteristic length scale and the behaviour of the
scaling function was found in both cases. We showed analytically within the mean field
approximation the transition from fractal to homogeneous distribution of persistent spins 
as the persistence exponent crosses $\frac{1}{2}$. We support our results by numerical simulations
in the kinetic $q$-state Potts model.

\section{Acknowledgments}
This research was supported in part by a grant (DMR 0088451) from the U.S National Science Foundation. The author
would like to thank P. Ray for a critical reading of the manuscript and suggestions for improvement.

\end{multicols}

\end{document}